\documentclass[twocolumn,  
 nofootinbib,   
 showpacs,  
groupedaddress,  preprintnumbers,  aps,  prd]{revtex4-1}
\usepackage{amsmath,  amssymb,   amsthm,  mathrsfs}
\usepackage{color}
\def\be{\begin{equation}}
\def\ee{\end{equation}}
\def\bea{\begin{eqnarray}}
\def\eea{\end{eqnarray}}
\def\ba{\begin{array}}
\def\ea{\end{array}}
\newcommand{\bes}{\begin{subequations}}
\newcommand{\ees}{\end{subequations}}

\begin{document}

\preprint{}
\title{Holographic Complexity}

\author{Mohsen Alishahiha}

\affiliation{School of physics,   Institute for Research in Fundamental Sciences (IPM)\\
P.O. Box 19395-5531,   Tehran,   Iran\\ 
{\rm email:  alishah@ipm.ir}}

\begin{abstract}
For a field theory with a gravitational dual,   following Susskind's proposal we define
holographic  complexity for a subsystem. The holographic  complexity 
is proportional  to the volume of a co-dimension one time slice  in the bulk geometry 
enclosed by the extremal
co-dimension two hyper-surface appearing in the computation of the holographic 
entanglement entropy. The proportionally constant,   up to a numerical order of one  factor 
is $G R$ where $G$ is the  Newton constant and $R$ is the curvature of the space time. 
We study this quantity in certain holographic {models}. We also explore a possible relation between  
the defined quantity and  fidelity appearing in  quantum information literature.
\end{abstract}


\maketitle

\section{Introduction}\label{intro}

AdS/CFT correspondence \cite{Maldacena:1997re} as a  concrete realization of holographic
principle \cite{{'tHooft:1993gx},  {Susskind:1994vu}} could provide a framework to
study quantum gravity and black hole physics. On the other hand  theoretical
quantum information may also provide a useful tool to study  physics of 
black holes in  gravitational theories. Therefore it would be interesting or might even 
be crucial to understand quantum information holographically,   in the sense that for
any quantity in  quantum information one could have a holographic dual description.

Holographic entanglement entropy \cite{Ryu:2006bv} is an explicit example in this paradigm which
is found useful to study quantum entanglement that might be eventually used to understand
the nature of the space time geometry. In quantum information there are other quantities,   
such as $n$-partite information, which might be of interest from holography point of view.  
Actually $n$-partite information  has been also  studied holographically in recent years
(see {\it e.g.} \cite{Alishahiha:2014jxa}).

We note, however, that  even if we could compute 
entanglement entropy or in general $n$-partite information (directly or holographically), 
 it might not be enough to
fully understand the system under consideration quantum mechanically.
It is  because no matter which entanglement
measure is being computed,  we may lose some information of the system simply because
the whole system is not a sum of the subsystems.
  
Actually it was recently pointed out that in order to understand properties of black hole horizons,  
it is also essential to consider quantum complexity\cite{Susskind:2014rva}. 
Suppose our system is in a given state and
we would like to map it to another state.  Then, intuitively,  the complexity tells us that 
how difficult
this task is. In fact,  it was conjectured that
for  an eternal black hole the complexity is proportional to
the spatial volume of the Einstein-Rosen bridge connecting  two boundaries 
\cite{Stanford:2014jda}.
 
Motivated by holographic entanglement entropy and quantum complexity, in this paper 
we would like  to further explore  a holographic description of {\it complexity} within
the context of AdS/CFT correspondence. \\
To proceed,  let us consider a field theory whose
holographic dual may be  provided by an Einstein  gravity  on  an asymptotically AdS geometry.
In this context  holographic entanglement entropy for a subsystem in the dual field theory can be
computed by  minimizing the area of a co-dimension two hyper-surface in the bulk geometry.

More precisely,  consider a subsystem $A$ in a time slice in the boundary theory. 
There is a minimal co-dimension two hyper-surface in the bulk,  denoted by $\gamma(A)$,  
whose boundary coincides  with the boundary of the subsystem $\partial\gamma=\partial A$.  
Then the holographic entanglement entropy is the area of the minimal surface divided by 
the Newton constant\cite{Ryu:2006bv}
\be
S_{\rm EE}=\frac{{\rm Area(\gamma)}}{4G}.
\ee

Based on this prescription here is an observation. Indeed the way
the  holographic entanglement entropy is computed would naturally
define,  rather uniquely, another quantity on the gravity side. Actually the minimal
hyper-surface considered above divides a constant time slice 
into two parts,  whose  {\it regularized} volumes are fixed as soon as the minimal hyper-surface
is determined. Therefore beside its area,   
the volume enclosed by the minimal hyper-surface may also define a new quantity.
 
To be  precise,  for a subsystem $A$ in the boundary theory,   let us denote by $V(\gamma)$
the volume of the part in the bulk geometry enclosed by the minimal hyper-surface
appearing in the computation of entanglement entropy. The corresponding part also
contains the subsystem $A$ itself.
Then motivated by  \cite{Susskind:2014rva} one may define {\it holographic  complexity}
as follows
\be\label{C}
{\cal C}_A=\frac{V(\gamma)}{8\pi R G },  
\ee
where $R$ is the radius of the curvature of the space-time,   {\it e.g.} AdS radius. The numerical factor
$8\pi$ is just a conventional factor. Clearly the definition is ambiguous up to an order of one 
numerical factor. It is also inherently  divergent and should be regularized by a UV cut off. 
Note also that by definition 
\be\label{INEQ}
{\cal C}_A+{\cal C}_{\bar{ A}}\leq  V_{ts},  
\ee
where  $\bar{A}$ is the complement of $A$ and $ V_{ts}$ is the whole regularized 
volume of the time slice. This inequality saturates for the ground state\footnote{
Note that one could also define another quantity in terms of the volume as 
${\cal B}_A=\frac{{\rm Max}\{V_A,  V_{\bar{A}}\}}{8\pi RG}$,   which for the ground state
one has ${\cal B}_A={\cal B}_{\bar{ A}}$.  Here $V_A$ ($V_{\bar A}$) is the volume in the bulk
associated to the subsystem $A$ (${\bar A}$).
Of course this is not the quantity we will consider in this 
paper. I would like to thank K. Papadodimas for a comment on this point.}.

Here we have implicitly assumed that the background is static,   though there is a natural 
generalization for time dependent geometries. Indeed,   following the covariant conjecture of 
holographic entanglement entropy \cite{Hubeny:2007xt} in order to define 
the corresponding holographic  complexity one should compute the 
volume of a part of  the space time 
enclosed by the  extremal co-dimension two hyper-surface appearing 
in the computation of the covariant holographic entanglement entropy\footnote{
Using the holographic description of mutual information,   one could obviously 
generalize the  holographic
complexity for multi subsystems.}.

It is worth noting that in the context of entanglement renormalization \cite{Vidal:2007hda}
the entanglement entropy may be estimated by the minimum number of bonds cut along 
a curve \cite{Swingle:2009bg} which could be thought of as Ryu-Takayanagi (RT) curve. Therefore,   
based on our definition of \eqref{C},  in this context the holographic complexity may be 
related (or estimated by) to the number of nods in the area enclosed by the 
curve cutting the bonds. Such a relation has also been suggested in 
\cite{Stanford:2014jda}. Using a holographic model,   it might be possible to make 
this statement more  precise\cite{TA}.

The aim of this paper is to  examine the quantity defined in the equation \eqref{C} for
 a certain holographic model. To be concrete we will consider a $d+1$ dimensional 
 CFT in its ground state whose 
dual description is given by a gravity on an $AdS_{d+2}$ geometry. By making use of the 
 gravity dual we will compute the holographic complexity. It is then natural to look for a proper
quantity in the field theory,   or in quantum information literature,    which could be 
identified as a holographic dual of the holographic complexity. 
Actually we will argue  that fidelity defined in  quantum information might  
provide such a dual quantity.

The paper is organized as follows. In the next section using an AdS geometry 
we will compute holographic complexity for a subsystem 
in the form of a sphere in a strongly coupled CFT. In section three we will compare the results 
obtained in the  section two
with fidelity defined for two states in a  CFT. The last section is devoted to discussions.


\section{Holographic computations}

Consider a gravitational theory on 
 an AdS$_{d+2}$ geometry which could provide a holographic dual 
for a $d+1$ dimensional  strongly coupled CFT in its ground state. Using RT prescription
\cite{Ryu:2006bv} one may compute holographic 
entanglement entropy for a sphere with radius $\ell$. To do so,   it is more convenient to 
take the following parametrization for the AdS geometry
\be
ds^2=\frac{R^2}{r^2}\left(-dt^2+dr^2+d\rho^2+\rho^2 d\Omega_{d-1}^2\right).
\ee
Then the entangling region is given by $t={\rm fixed},  \;\rho\leq \ell$. 
To compute the holographic entanglement
entropy  one needs to minimize the area of a co-dimension two hyper-surface in 
the bulk which may be parametrized by $\rho=f(r)$. It is easy to see that the area  
is minimized for $f(r)=\sqrt{\ell^2-r^2}$\cite{Ryu:2006bv}. 
 
Following our proposal one needs to 
evaluate  the volume enclosed by the
above minimal area
\bea
V&=&\Omega_{d-1}R^{d+1}\int_{\rho\leq f(r)} d\rho \;dr\;\frac{\rho^{d-1}}{r^{d+1}}\\
&=&
\frac{\Omega_{d-1}R^{d+1}}{d}\int_\varepsilon^\ell dr\;\frac{(\ell^2-r^2)^{d/2}}{r^{d+1}},  
\nonumber
\eea
where $\Omega_{d-1}$ is the volume of the unit sphere  $S^{d-1}$ and
$\varepsilon$ should be thought of as a UV cut off. It is easy to perform the integration
over $r$ to find the holographic  complexity. Indeed 
for even dimensional CFT's  (odd  $d$ in our notation) one arrives at
\bea\label{Codd}
{\cal C}_A\!&\!=\!&\!\frac{\Omega_{d-1} R^{d}}{8d \pi G}\bigg(\frac{1}{d} \frac{\ell^{d}}{\varepsilon^d}-
\frac{d}{2(d-2)}\,   \frac{\ell^{d-2}}{\varepsilon^{d-2}}+\frac{d(d-2)}{8(d-4)}\,   \frac{\ell^{d-4}}{\varepsilon^{d-4}}\nonumber \\ && \;\;\;\;\;\;\;\;\;\;\;\;\;\;\;\;
+\cdots -(-1)^{[\frac{d}{2}]}\;\frac{\pi}{2}\bigg),  
\eea
for $d=1,  3,  5,  \cdots$. Here $[y]$ denotes the integer part of $y$. 
On the other hand for  odd dimensional CFT's (even $d$ ) one gets 
\bea
{\cal C}_A&\!=\!&\frac{\Omega_{1} R^{2}}{16\pi G}\left( \frac{\ell^{2}}{2\varepsilon^2}-\log\frac{\ell}{\varepsilon}
-\frac{1}{2}\right),  \\ 
{\cal C}_A&\!=\!&\frac{\Omega_{3} R^{4}}{32\pi G}\left( \frac{\ell^{4}}{4\varepsilon^4}
-\frac{\ell^2}{\varepsilon^2}
+\log\frac{\ell}{\varepsilon}
+\frac{3}{4}\right),  \nonumber\\
{\cal C}_A&\!=\!&\frac{\Omega_{5} R^{6}}{48\pi G}\left( \frac{\ell^{6}}{6\varepsilon^6}
-\frac{3\ell^4}{4\varepsilon^4}+\frac{3\ell^2}{2\varepsilon^2}-
\log\frac{\ell}{\varepsilon}
-\frac{11}{12}\right),  \nonumber
\eea 
for $d=2,  4,  6$,   respectively. It is interesting to note that for odd dimensions 
the holographic complexity contains a  logarithmic divergent term.

It is also worth noting that the most divergent term in the expression of holographic 
 complexity is proportional to the volume of the subsystem $V(A)$
\be
{\cal C}_A=\frac{R^d}{8d \pi G}\;\frac{V(A)}{\varepsilon^d}+\cdots,  
\ee
leading to a volume law behavior. This should be thought of as an analogous 
to the celebrated area law of the entanglement entropy. Moreover for arbitrary $d$, 
the holographic  complexity contains a universal term  in the sense that it is  independent of 
the UV cut off. For odd $d$ the universal term can be  identified with
the finite term, while  for even $d$ it  is given by  the coefficient of the log divergence term
   \bea\label{G2}
{\cal C}_A^{\rm uni}=(-1)^{[\frac{d}{2}]}\Bigg\{ \begin{array}{rcl}
&\frac{\Omega_{d-1} R^{d}}{16d  G}&\,  \,  \,  {\rm odd}\,  d,  \\ & &\\
&\frac{\Omega_{d-1} R^{d}}{8d \pi G}&\,  \,  \,  {\rm even}\,  d\,. 
\end{array}
\eea
Note that the universal terms are  also independent of the size of  the subsystem $\ell$,  
 indicating that it could reflect certain intrinsic properties of the theory under consideration. 
 In fact  it might be thought of as a central charge for the model.

It is also interesting to compute the holographic complexity for a subsystem in an excited state. 
Holographically an excited state may be described by an asymptotically AdS geometry.
To be concrete let us consider an AdS black hole whose metric,   adopted to our purpose,    may be 
written as follows
\be
ds^2=\frac{R^2}{r^2}\left(-h(r)  dt^2+\frac{dr^2}{h(r)}+d\rho^2+\rho^2 d\Omega_{d-1}^2\right),  
\ee
where $h(r)=1-m r^{d+1}$ with $m$ being a constant. In this case,   similar to the  
pure AdS background,   the minimal hyper-surface in the 
bulk geometry \eqref{G2} associated with a sphere subsystem may be parametrized by $\rho=f(r)$,   
though in the present case the function $f$ does not have a closed simple form. 
Nonetheless one can find the profile of the minimal surface at leading 
order in $m$. More precisely assuming $m\ell^{d+1}\ll 1$ 
one finds \cite{Bhattacharya:2012mi} 
\be
f(r)=\sqrt{r_t^2-r^2}\bigg(1+\frac{2 r_t^{d+3}-r^{d+1}(r_t^2+r^2)}{2(d+2)(r_t^2-r^2)}m\bigg)+{\cal
O}(m^2),  
\ee
where the turning point $r_t$ at leading order in $m$ is given by
\be\label{tp}
r_t=\ell\left(1-\frac{m\ell^{d+1}}{d+2}+{\cal O}\bigg((m\ell^{d+1})^2\bigg)\right).
\ee
The volume enclosed by the minimal area is
\be
V=\frac{\Omega_{d-1} R^{d+1}}{d}\int_\varepsilon^{r_t} dr\,  \frac{f(r)^d}{r^{d+1}\sqrt{h(r)}}.
\ee
The above integral can  be evaluated order by order in $m$ and  the result would be 
a function of the turning point $r_t$. On the other hand using the expression of 
the turning point \eqref{tp}  one can re-write  the holographic complexity as a 
function of the radius  $\ell$ at leading order in $m$. Doing so,  in the present case   
unlike the entanglement entropy,   
one finds that at leading order the holographic  complexity remains uncharged
\be
{\cal C}_A^{\rm BH}={\cal C}_A^{\rm AdS}+{\cal O}(m^2).
\ee
In fact one could go further to evaluate the order of $m^2$ term as well. 
Although the expressions  are lengthy,   the final result is simple,   given by
\be
\Delta {\cal C}_A=  {\cal C}_A^{\rm BH}-{\cal C}_A^{\rm AdS}=
c_d\frac{\Omega_{d-1} R^{d}}{8d \pi G} 
\left(m\ell^{d+1}\right)^2,  
\ee
where $c_d$ is a calculable (non-negative) numerical constant. For example one has 
$c_1=0,  c_2=\frac{1}{128},  \cdots$.

This result may be compared with that of the entanglement entropy where one finds 
that the variation of entanglement entropy gets corrected at order of $m\ell^{d+1}$
\cite{{Bhattacharya:2012mi},  {Allahbakhshi:2013rda}}
\be
\Delta S_{\rm EE}=S_{\rm EE}^{\rm BH}-S_{\rm EE}^{\rm AdS}=
{\tilde c}_d \frac{\Omega_{d-1} R^d}{G}\left(m\ell^{d+1}\right),  
\ee
leading to the first law of entanglement\cite{Blanco:2013joa}. Here ${\tilde c}_d $ is a numerical factor
(see {\it e.g.}  \cite{Bhattacharya:2012mi}). Note that $\epsilon=m\ell^{d+1}$ is the 
expanding parameter which measures  how much the  ground state is deformed to  the excited state. 
Taking into account that in the present case the energy of the excited state is proportional to 
${\cal E}\sim \frac{R^d}{G}ml^d$, up to numerical factors, one arrives at
\bea
\Delta S_{\rm EE}&\sim&\frac{R^d}{G} \epsilon\sim {\cal E}\ell,  \\ \nonumber
\Delta {\cal C}_A&\sim& (d-1)
\frac{R^d}{G}\epsilon^2\sim (d-1)
\frac{G}{R^d}{\cal E}^2 \ell^2,  
\eea
{\it i.e.} while the change of entanglement entropy gets first order correction with respect to 
energy, the holographic complexity receives second order correction. Here we have explicitly put the factor 
of $d-1$ to stress that for $d=1$ the correction vanishes.

Finally we should note that although we have done all the computations for a subsystem  with 
spherical symmetry,   it could be done for other subsystem  such as a strip. An advantage to work 
with sphere is that we could present the results analytically.

 \section{Fidelity and holographic complexity}

The complexity as defined in \cite{Susskind:2014rva} is a quantity  to measure how
difficult a task is. Given a quantum system a task would be a unitary evolution to  map a state 
to another state. In quantum information  there are several quantities which could 
provide measures to compare two states. These  include,   for example,    relative entropy or
fidelity (see \cite{Lin:2014hva} for a holographic description of the relative entropy).  
The aim of this section is to investigate  whether there is any connection 
between holographic  complexity and fidelity.

To explore this  point let us start with  a quantum pure state  $|\Psi(\lambda_1)\rangle$
 in the Hilbert of a quantum system. Where $\lambda$ is a tunable parameter of
the model. Now consider a neighboring pure 
 state $|\Psi(\lambda_2)\rangle$ which may be reached by  changing,   
 infinitesimally,   the  parameter $\lambda$.  It is then natural to pose a question  how 
 close these two states are? To address this question one could compute fidelity 
 \cite{Uhlmann:1976} which in the present case where both states are pure  it is given by the inner 
 product of the two states. For sufficiently small perturbation $\delta\lambda=\lambda_2-\lambda_1$
 one has
\be \label{P}
|\langle \Psi(\lambda_1) |\Psi(\lambda_2)\rangle|=1-G_\lambda\; \delta\lambda^2+{\cal O}(\delta^3),  
\ee
where   $G_\lambda$ is fidelity  susceptibility.    This expression,   considered as a 
 metric (see {\it e.g.} \cite{Braunstein:1994zz}),    could  measure the distance between two 
 neighboring quantum pure states.

Recently a gravity dual  for information metric was proposed in \cite{MIyaji:2015mia}
where it was suggested  that under certain approximations the fidelity susceptibility 
for a $d+1$ dimensional 
CFT deformed by a marginal perturbation is holographically given by 
the time slice with the maximal volume in the AdS background which ends on the time slice
 at the AdS boundary \cite{MIyaji:2015mia}.

From our construction it is clear that within the same approximations if
one takes infinite volume limit, the holographic complexity reduces to the 
fidelity susceptibility  $G_\lambda$ as computed in \cite{MIyaji:2015mia}.
In particular for a two dimensional CFT from  \eqref{Codd} one has 
\be\label{C2}
{\cal C}_A=\frac{c\ell}{12\pi \varepsilon}-\frac{c}{24},  
\ee
 where $c=\frac{3R}{2G}$ is the central charge of the two dimensional
CFT. Clearly in the large $\ell$ limit it reduces to  
that in \cite{MIyaji:2015mia} obtained from AdS Janus solution 
\cite{Bak:2007jm}. It is interesting to note that the finite term in the above 
expression is proportional to the Casimir energy of the two dimensional CFT which typically 
appears whenever we are dealing with a CFT in a finite volume. Note that 
the factor of 24 comes from our particular normalization of ${\cal C}_A$.

More generally for a $d+1$ dimensional CFT one gets
\be
{\cal C}_A|_{\ell \rightarrow \infty}= \frac{V R^d}{8\pi d G \varepsilon^d},  
\ee
where $V$ is the volume of the  time slice in the AdS geometry. This is exactly the
fidelity  susceptibility obtained in \cite{MIyaji:2015mia} for an AdS background with a
defect  brane\cite{Karch:2000ct} considered as a marginal deformation.
It is important to note that to get the right $\frac{1}{\varepsilon^d}$ behavior it is crucial 
to deform the CFT by an exactly marginal  operator \cite{MIyaji:2015mia}. 
In terms of the information metric $G_\lambda$ the inequality \eqref{INEQ} reads
${\cal C}_A+{\cal C}_{\bar A}\leq G_\lambda$.

Note also that the  above comparison  works just for the extremely large $\ell$ limit. 
In other words there is,   a priori,   no way to understand 
the subleading  divergences in this picture,   nor it is not clear how to compare the finite temperature 
case where the  temperature dependent term drops in the large $\ell$ limit.

Therefore although this comparison seems reasonable,   it is not quite clear 
to us whether there is a relation between holographic  information metric obtained 
in \cite{MIyaji:2015mia} and the holographic  complexity studied in this paper (or that 
defined in \cite{Susskind:2014rva})\footnote{I would like to thank T. Takayanagi for a 
comment on this point.}.
 Nonetheless it might be possible to extend 
the notation of  fidelity  susceptibility for a subsystem with a finite size.

To explore this point better it is useful to write the fidelity in terms of 
a density metric. Denoting by  $\varrho(\lambda_1)$ and $\varrho(\lambda_2)$ the 
density matrices associated with two states $|\Psi(\lambda_1)\rangle$ and 
 $|\Psi(\lambda_2)\rangle$ respectively,    the fidelity can be written as follows \cite{Uhlmann:1976}
\be
F={\rm Tr} \sqrt{\sqrt{\varrho(\lambda_1)}\;\varrho(\lambda_2)\;\sqrt{\varrho(\lambda_1)}},  
\ee
which for pure states  reduces to the inner product of the states,  
$|\langle \Psi(\lambda_1) |\Psi(\lambda_2)\rangle|$.  On the other hand dealing with  a subsystem, 
it is natural to consider a reduced density matrix and 
therefore define the fidelity for two reduced density matrices \cite{Zhou:2007}
\be
F_A={\rm Tr}_A\sqrt{\sqrt{\varrho_A(\lambda_1)}\;\varrho_A(\lambda_2)\;\sqrt{\varrho_A(\lambda_1)}},  
\ee
where  $\varrho_A(\lambda_1)$ and $\varrho_A(\lambda_2)$ are the corresponding reduced 
density matrices. Now the aim is to expand the reduced fidelity 
for a small perturbation to find an expression for {\it reduced fidelity susceptibility}.

To  proceed, we will take  advantage of having   a subsystem in the shape of a sphere.
 Actually  when we have a subsystem with spherical symmetry in  the ground state of a CFT,   one may conformally map the system to a  thermal system whose temperature is given by the radius of the sphere; $\beta=2\pi \ell$ \cite{Casini:2011kv}.
  More precisely,   under this conformal map
 the reduced density matrix  maps to a thermal density matrix given by \cite{Casini:2011kv}
  \be
 \varrho_{\rm th}(\lambda)=\frac{e^{-2\pi \ell H_\tau (\lambda)}}{{\rm Tr}(e^{-2\pi \ell H_\tau (\lambda)})},  
 \ee
 where $H_\tau$ is the standard Hamiltonian of the thermal system which 
 corresponds to the time translation. As a result we will have to compute fidelity at 
  finite temperature. Fidelity for a mixed state at finite temperature has been studied 
 in \cite{Zanardi:2007} where the authors have considered two possibilities; 
 either to change the temperature while keeping the parameter $\lambda$ fixed, or 
 another way around. In our model since  
 we are dealing with a system with a fixed temperature (fixed $\ell$) the corresponding 
 thermal fidelity should be given as follows \cite{Zanardi:2007} 
 \be
F(2\pi\ell,  \lambda_1,  \lambda_2)={\rm Tr}\sqrt{\sqrt{\varrho_{\rm th}(\lambda_1)}\;\varrho_{\rm th}(\lambda_2)\;\sqrt{\varrho_{\rm th}(\lambda_1)}}.
\ee
Setting $\lambda_2=\lambda_1+\delta\lambda$,   for sufficiently small $\delta\lambda$ and using 
the definition of thermal density matrix in terms of the Hamiltonian one gets \cite{Quan:2008}
\be
F(2\pi\ell,  \lambda_1,  \lambda_2)= 1-2\pi \ell \; \chi_\lambda \;
\frac{\delta\lambda^2}{8}+{\cal O}(\delta\lambda^3),  
\ee
where $\chi_\lambda=\partial^2_\lambda {{\cal F}_{\rm th}}$ is the
 fidelity susceptibility given in terms
of  the free energy of the thermal system, ${{\cal F}_{\rm th}}$. 
If one perturbs the system by an operator 
with dimension $\Delta$,   then the fidelity susceptibility scales as $\frac{R^d}
{\varepsilon^{2\Delta-2-d}}$, where $R$ is a scale of the model (see {\it e.g.} \cite{Gu:2008}).
 Thus for a marginal operator where 
$\Delta=d+1$  one gets $\chi_\lambda\sim (R/\varepsilon)^d$. 
On the other hand the free energy and 
therefore  susceptibilities  receive finite temperature corrections which have an  expansion
 in power of
$T^2$ (see {e.g. \cite{Laine}). Therefore for our mixed thermal  state, 
where the temperature is given by $T=\frac{1}{2\pi\ell}$,  
one arrives at 
\be
\chi_\lambda\sim \frac{R^d}{\varepsilon^d}\left(1+c_2 \frac{\varepsilon^2}{\ell^2}+
c_4 \frac{\varepsilon^4}{\ell^4}+\cdots\right),  
\ee
in qualitative agreement with our results in the previous section.

It is also interesting to use the inverse of the conformal map to return to the original picture
of the reduced density matrix. In fact doing so,   one gets
\be
F_A=1-\partial_\lambda^2 {\cal F}\; \frac{\delta\lambda^2}{8}+{\cal O}(\delta\lambda^3),  
\ee
where ${\cal F(\lambda)}={\rm Tr}(\varrho_A(\lambda) H(\lambda))-S_{\rm EE}(\lambda)$ 
with $S_{\rm EE}$  being the entanglement entropy and,    $H(\lambda)$ is the modular
 Hamiltonian by which the  reduced density matrix may be given as follows 
\be
\varrho_A(\lambda)=\frac{e^{-H(\lambda)}}{{\rm Tr}(e^{-H(\lambda)})}.
\ee
Note that in terms of the modular Hamiltonian the fidelity susceptibility 
$\chi=\partial_\lambda^2 {\cal F}$ is given by
\be
\chi=\langle H^2\rangle-\langle H\rangle ^2.
\ee
Since the explicit form of the modular Hamiltonian 
is known for the subsystem we are considering\cite{Blanco:2013joa}, it would be interesting to find the fidelity susceptibility
directly from the modular Hamiltonian.  We are currently working on this line.

\section{Discussions}
In this paper we have defined and studied holographic complexity for a 
subsystem in a CFT which has a holographic description. Motivated by  
the holographic entanglement entropy,   the  corresponding quantity has been 
 defined by the volume enclosed by the extremal co-dimension
two hyper-surface appearing in the computation of the holographic entanglement entropy.

We have also compared the  holographic complexity of the ground state
of a CFT deformed by a marginal operator with the reduced fidelity susceptibility where we have seen 
that these two quantities qualitatively behave in a similar manner. 
We have also noticed that in large volume limit the holographic 
complexity reduces to holographic information metric studied in  \cite{MIyaji:2015mia}.
 
It is also  illustrative to apply the above suggestion for the case of  thermofield doubled CFT's whose
gravitation dual may be provided by an eternal black hole. The entanglement entropy for this
model has been studied in \cite{Hartman:2013qma} where it was shown that the holographic
entanglement entropy is given by the area of an extremal surface connecting two boundaries. 
Following the equation \eqref{C}, one needs to compute the volume enclosed by this 
extremal hyper-surface. Actually for this system the complexity has been studied in
\cite{Stanford:2014jda} where it was shown that the complexity is given by the spatial 
volume of the Einstein-Rosen bridge connecting  two boundaries.

It is worth mentioning that through the whole of this paper we have implicitly assumed 
that the gravitational theory is given by an Einstein gravity. It is then natural to ask how to define 
 holographic complexity when we have a gravitational theory with higher derivative terms?  
 Note that for this case the thermal entropy and holographic entanglement entropy are given by
 Wald entropy and  the generalized entropy functional (see {\it e.g.} \cite{Dong:2013qoa}),  
 respectively. Therefore we would expect to have a general expression (which is
 not necessary the volume) for the holographic complexity as well. 

Actually to address this question one needs to understand two issues. First of all one 
should understand  how to specify the corresponding co-dimension two
hyper-surface in the bulk.  And
secondly, even when we have fixed a part of the time slice,   what quantity should be evaluated 
in this  time slice?

In fact  there is  a natural proposal  for  holographic complexity in a general 
gravitational theory which is as follows. Consider the Wald charge
appearing as the integrand in the Wald formula for entropy. 
Then evaluate it on a co-dimension one time slice enclosed by a co-dimension
 two hyper-surface minimizing the entropy functional appearing in the computation of holographic entanglement entropy. More explicitly 
\be
{\cal C}_A=-\frac{1}{R}\int_{\rm enclosed\,   volume} \;\;\frac{\partial {\cal L}}{\partial R_{\mu\nu\rho\sigma}}\epsilon^{\mu\nu}\epsilon^{\rho\sigma},  
\ee \\
where $ \epsilon^{\mu\nu}$ is binormal to the co-dimension two hyper-surface enclosing 
the volume.   Here the factor of  $\frac{1}{R}$   comes form a dimensional analysis and 
the fact that the only 
natural dimensionfull parameter of the model is the curvature radius of the space time. 
In fact this factor is the same as the extra $R$ appearing in  equation \eqref{C}.
 Clearly for Einstein gravity the above expression  reduces  to the volume. On the other hand if  
 one computes the holographic complexity for a sphere in the most general quadratic action
 \be
 \int d^{d+2}x\;\sqrt{g}\;(\alpha_1 R_{\mu\nu\rho\sigma}  R^{\mu\nu\rho\sigma}  
 +\alpha_2 R_{\mu\nu} R^{\mu\nu}+\alpha_3 R^2),  
 \ee
 then it gets corrected by an overall factor  given by $4\alpha_1+2(d+1)(\alpha_2+(d+2)\alpha_3)$. 
 Interestingly enough it is exactly the factor 
 which appears in the correction of the entanglement entropy
 for a sphere. It is then natural to think of the  universal term of holographic complexity 
 as the central charge of the model (see also \eqref{C2}).

Clearly this point deserves more investigations. We would like to mention that quantum complexity
for a general theory is  extensively studied  in \cite{coming}. Of course it is not clear to us 
whether,   for a generic case,    there is a 
direct relation between quantum complexity studied in \cite{coming} and 
holographic complexity studied in the present paper.

\begin{acknowledgments}

I would like to thank Luis Alvarez-Gaume and  Kyriakos Papadodimas for  discussions. 
The author would also like to thank  Vahid Karimipour,   Leonard Susskind and Tadashi Takayanagi
 for correspondences  and comments. 
 Special thank to Leonard Susskind for encouragements and
 providing me a draft of \cite{coming} prior to publication. I would also like to thank CERN 
 TH-Division for very warm hospitality. This work is supported by Iranian National Science
 Fundation (INSF).

\end{acknowledgments}

\end{document}